\pdfoutput=1
\documentclass[cameraready]{Interspeech}
\title{GigaAM Multilingual: Foundation Model for Underrepresented Languages}

\author[affiliation={1}, equalcontribution]{Andrei}{Kuzmenko}
\author[affiliation={1}, equalcontribution]{Alexandr}{Maximenko}
\author[affiliation={1}]{Aleksandr}{Kutsakov}
\author[affiliation={1}, correspondingauthor]{Georgii}{Gospodinov}
\author[affiliation={1}]{\\Dmitrii}{Bolotov}
\author[affiliation={1}]{Oleg}{Kutuzov}
\author[affiliation={1}]{Pavel}{Bogomolov}
\author[affiliation={1}]{Fyodor}{Minkin}

\address{
    $^1$ SaluteDevices, Russia
}

\email{\{andrey.kuzmenko2907, ae.maximenko, askutsakov, georgygospodinov, bolotovdm, olegkutuzov01, bobrosoft98, minkin.fyodor\}@gmail.com}


\usepackage{booktabs}
\usepackage{multirow}

\begin{document}

\maketitle

\begin{abstract}
    Despite recent scaling successes, multilingual ASR performance remains highly uneven, with long-tail languages suffering from severe data scarcity. This work addresses the challenge of building robust foundation models for underrepresented Central Asian languages (Kazakh, Kyrgyz, Uzbek). We present GigaAM Multilingual, a Conformer encoder pre-trained on 2M hours of audio using a HuBERT-style objective. Crucially, we introduce a cluster-level data balancing strategy during pre-training and a domain-aware sampling method during fine-tuning to mitigate head-language dominance. In controlled comparisons, our approach outperforms strong open pretrained encoders (Whisper Large v3, Omnilingual-1B) on target languages, achieving significant gains on spontaneous speech while maintaining efficiency. We release the foundation encoder and ASR model, offering a proven recipe for effective multilingual adaptation under realistic data imbalance.
\end{abstract}

\section{Introduction}

Recent years have seen rapid progress in multilingual automatic speech recognition (ASR), driven by scaling data, model capacity, and weakly-supervised and self-supervised training objectives \cite{whisper,usm,mms,omnilingual}. Despite these advances, recognition quality remains highly uneven across languages: performance on high-resource languages is strong, while many low-resource and long-tail languages still exhibit error rates that are prohibitive for downstream applications \cite{ml_superb_2}.

A major contributor to this disparity is data imbalance \cite{mhubert147}: empirical data distribution favors head languages, whereas drastic equalization via naive upsampling can lead to overfitting or degradation on high-resource languages. Principled strategies for mixing and weighting data in pre-training and adaptation are thus essential.

In this work, we extend GigaAM \cite{gigaam} -- an efficient self-supervised learner (SSL) for Russian ASR -- to the multilingual setting. We pre-train the encoder on a 2M-hour corpus with highly skewed language-group proportions (Fig.~\ref{fig:lang_distribution}). Although the target Central Asian languages (Kyrgyz, Kazakh, Uzbek) are present in pre-training, their share is small, and strong general-purpose ASR models often struggle in this low-resource regime (Table~\ref{tab:model_performance}).

\begin{table}[t]
\caption{WER (\%) of GigaAM Multilingual against public multilingual ASR systems on Common Voice (CV), FLEURS, and our internal in-the-wild test sets (Sec.~\ref{evaluation}), best in bold.}
\label{tab:model_performance}
\centering
\setlength{\tabcolsep}{4pt}
\resizebox{\linewidth}{!}{%
\begin{tabular}{llcccc}
\toprule
Lang & Split & \textbf{\shortstack{GigaAM\\ Multilingual}} & \textbf{\shortstack{Omnilingual\\ 1B LLM ASR}} & \textbf{\shortstack{Seamless M4T\\ large v2}} & \textbf{\shortstack{Whisper\\large v3}} \\
\midrule
\multirow{2}{*}{English} & CV     & 21.5 & 24.7 & \textbf{16.2} & 20.0 \\
                        & FLEURS &  9.4 &  7.1 &  5.8 & \textbf{3.9} \\
\midrule
\multirow{3}{*}{Russian} & CV     & \textbf{5.1} & 13.6 &  9.2 &  9.1 \\
                         & FLEURS & \textbf{3.0} &  6.4 &  4.6 &  3.1 \\
                         & Internal    & \textbf{6.0} & 14.6 & 16.1 & 10.1 \\
\midrule
\multirow{3}{*}{Kazakh} & CV     & \textbf{13.8} & 23.7 & 23.8 & 57.8 \\
                        & FLEURS & \textbf{4.4}  &  6.6 &  6.8 & 32.4 \\
                        & Internal    & \textbf{15.8} & 32.2 & 62.9 & 65.2 \\
\midrule
\multirow{3}{*}{Kyrgyz} & CV     & \textbf{10.2} & 21.6 & 14.3 & 95.2 \\
                        & FLEURS & \textbf{5.5}  &  8.1 &  9.5 & 86.3 \\
                        & Internal    & \textbf{9.8} & 25.0 & 78.3 & 102.2 \\
\midrule
\multirow{3}{*}{Uzbek} & CV     & \textbf{9.2} & 32.8 & 25.1 & 109.9 \\
                       & FLEURS & \textbf{7.3}  & 15.4 & 11.9 & 105.4 \\
                       & Internal    & \textbf{12.7} & 30.2 & 40.0 & 120.6 \\
\bottomrule
\end{tabular}}
\end{table}

We propose a two-level approach that explicitly accounts for long-tail structure. First, we incorporate group-aware reweighting during SSL pre-training, increasing the contribution of low-share language groups to the pre-training objective via cluster-level sampling weights. Second, we fine-tune on a mixture of Russian, English, Kyrgyz, Kazakh, and Uzbek using a data recipe that combines open-source corpora, synthetic speech, weakly-supervised and internally annotated data.

We ablate pre-training and fine-tuning sampling and evaluate adaptation to low-coverage languages. Under a controlled protocol with matched fine-tuning data and decoding, GigaAM adapts better than Whisper \cite{whisper} and Omnilingual \cite{omnilingual} pretrained encoders on Kyrgyz/Kazakh/Uzbek, and converges faster on languages with minimal pre-training coverage (e.g., Bashkir, Georgian).

We summarize our contributions as follows:
\begin{itemize}
\item GigaAM Multilingual: a state-of-the-art open-weight\footnote{\url{https://github.com/salute-developers/GigaAM}} multilingual ASR model for Kyrgyz, Kazakh, and Uzbek.
\item Group-aware reweighting during SSL pre-training and its interaction with downstream fine-tuning.
\item Cross-lingual transfer results on tail languages.
\end{itemize}

\section{Related Work}

Large-scale ASR models have substantially improved multilingual robustness by scaling training data and model capacity. Whisper demonstrates strong zero-shot generalization from large-scale weak supervision \cite{whisper}, while USM and MMS improve coverage and transfer via large-scale multilingual pre-training and downstream fine-tuning \cite{usm,mms}. Omnilingual ASR explicitly targets extensibility to a very large set of languages and emphasizes rapid adaptation with minimal data \cite{omnilingual}. Despite this progress, recent benchmark efforts highlight that improvements remain uneven across languages and varieties, with long-tail languages still lagging behind \cite{ml_superb_2}.

\subsection{Multilingual speech representation learning}

Self-supervised learning enables the training of transferable speech encoders from unlabeled audio \cite{wav2vec2,hubert,xlsr}. Multilingual SSL work shows that scaling alone is insufficient: sampling must account for language and dataset imbalance to prevent head languages from dominating \cite{mhubert147}. mHuBERT-147 scales HuBERT-style training to 147 languages with multilingual up-sampling and batching. This motivates our group-aware reweighting during pre-training and its interaction with downstream adaptation.

\subsection{Data pipelines for low-resource ASR}
For tail languages, the primary bottleneck is often data: transcribed speech is scarce and noisy. Consequently, recent efforts increasingly adopt automated corpus construction and refinement pipelines.

GigaSpeech 2 provides a concrete example of an end-to-end pipeline for creating a low-resource corpus, using Whisper for initial transcription, MMS-based forced alignment, and iterative refinement to improve pseudo-labels \cite{gigaspeech_2}. At the same time, data heterogeneity itself can be a dominant factor: OWSM v3.2 analyzes how heterogeneous sources affect speech-to-text foundation models and proposes strategies for filtering and normalization \cite{owsm3_2}, while OWSM v4 emphasizes scalable data cleaning and web-data integration \cite{owsm4}. These trends underscore that data curation and mixing policies are fundamental design factors; our work turns this insight into practice by quantifying the contribution of each pipeline component and by showing that source-aware weighting is critical for stable low-resource gains (Section \ref{data_ablation}).

\subsection{Practical encoder selection for low-resource adaptation}
While open multilingual encoders make low-resource adaptation accessible, selecting a suitable starting point is non-trivial due to undisclosed data mixtures and limited ablations.

Motivated by this gap, we perform a controlled comparison of two recent multilingual encoders trained at large scale (approximately 3--5M hours). We evaluate these encoders under an identical adaptation pipeline on our target language set, using the same fine-tuning recipe, weighting strategy, and decoding configuration (Section~\ref{encoder_ablation}).

\section{Method}

\subsection{Pre-training}

We pre-train GigaAM Multilingual with a HuBERT-style \cite{hubert} masked unit prediction objective, where the model predicts discrete acoustic units at masked time steps from an input mel-spectrogram sequence. Both the teacher and the student are 600M-parameter Conformer encoders with 24 layers, a hidden dimension of 1024, and Rotary Position Embeddings~\cite{su2021rope} in the self-attention layers. The encoder operates at a frame rate of 25\,Hz (40\,ms stride).

Target units are produced by the teacher model. We extract teacher representations on the training split and run $K$-means with $K{=}1000$ clusters over these features. Each representation is then mapped to its nearest centroid to obtain a discrete unit label.

In pre-training, we randomly mask contiguous spans covering 40\% of the input frames and train the student to predict the teacher-derived labels at the masked positions.

\subsection{Data Preparation and Weighting}

We construct a corpus of 2 million hours of in-house audio. For pre-training data preparation, we virtually segment each audio file into one-minute chunks and apply an in-house voice activity detection model to every chunk. As a result, each recording is reduced to several speech-only fragments. We then run language identification on every fragment using the MMS LID 4017 model \cite{mms}. Since a single audio may contain multiple fragments, we determine the language of the full recording via majority voting over fragment-level predictions. This procedure yields more than 70 high-frequency languages in the pre-training corpus.

Directly balancing all 70 languages is particularly challenging due to the presence of low-resource languages, for which reliable weight estimation is difficult and prone to overfitting. To address this issue, we apply a clustering algorithm \cite{cluster} to a weighted language co-occurrence graph. In this graph, vertices correspond to languages, and edge weights reflect how often a pair of languages appears within the same audio recording. Importantly, we do not use raw co-occurrence counts: we instead use a normalized weight $w_{ij} = c_{ij} / \sqrt{n_i n_j}$ (with $c_{ij}$ = co-occurrence count, $n_i$, $n_j$ = counts per language), so pairs co-occurring more than expected get higher weights.

\begin{figure*}[!t]
  \centering
  \includegraphics[width=\textwidth]{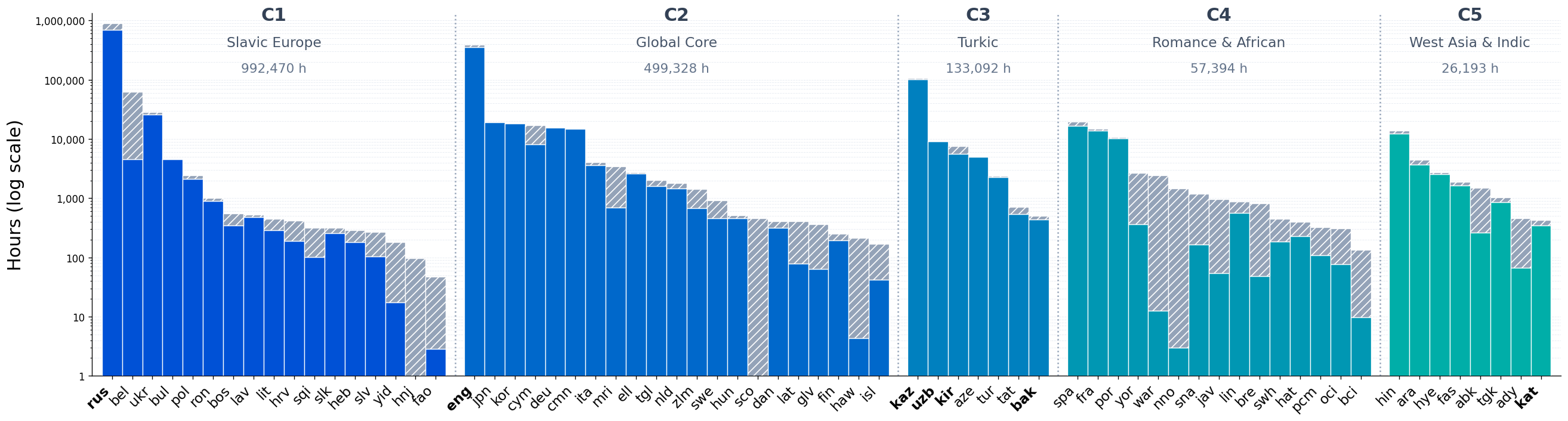}
  \caption{
  Language duration distribution by cluster. Bars show hours per language; shaded ``uncertain'' segments denote low-confidence predictions assigned to the corresponding language. Languages in bold participate in the experiments.
  }
  \label{fig:lang_distribution}
\end{figure*}

After pruning low-weight edges and low-degree vertices, clustering yields five language clusters (Fig.~\ref{fig:lang_distribution}); otherwise, the graph fragments into small outlier groups, making balancing during pre-training more difficult. Audio recordings are assigned to languages using two thresholds: 0.7 on majority-vote confidence and 0.7 on the mean model confidence of the selected-language segments. Audio recordings below either threshold are treated as ``uncertain'' in the corresponding language. Subsequent data weighting is performed at the cluster level rather than per individual language.

\subsection{Fine-tuning}

We fine-tune the pretrained encoder with a CTC objective \cite{ctc} using a shared character vocabulary across English, Russian, Kazakh, Kyrgyz, and Uzbek. To address data scarcity for Central Asian languages, we train on a multi-domain dataset comprising open-source, internal, synthetic, and weakly-supervised data (Table~\ref{tab:sft_distribution}).

\begin{table}[ht]
\caption{Fine-tuning data (hours) by language and source.}
\label{tab:sft_distribution}
\centering
\resizebox{\linewidth}{!}{%
\begin{tabular}{l c c c c c}
\toprule
 & \textbf{English} & \textbf{Russian} & \textbf{Kazakh} & \textbf{Kyrgyz} & \textbf{Uzbek} \\
\midrule
Open-Source & 26738 & 1767 & 1031 & 26 & 313 \\
Crowd-Source & 0 & 3363 & 876 & 514 & 272 \\
Weakly-Supervised & 0 & 0 & 9  & 324 & 0 \\
Synthetic & 0 & 692 & 7896 & 6415 & 0 \\
\bottomrule
\end{tabular}}
\end{table}

\subsubsection{Open-Source Data}

We incorporate publicly available corpora covering read and conversational speech across the target languages, including Common Voice \cite{common_voice}, FLEURS \cite{fleurs}, Golos \cite{golos}, Russian LibriSpeech \cite{rus_librispeech}, Europarl-ASR \cite{europarl_asr}, LibriSpeech \cite{librispeech}, People's Speech \cite{peoples_speech}, SLURP \cite{slurp}, SPGISpeech \cite{spgispeech}, TED-LIUM \cite{tedlium}, VCTK \cite{vctk}, VoxForge \cite{voxforge}, Uzbek Speech Corpus \cite{usc},  Kazakh Speech Corpus 2 \cite{ksc2}, KazakhTTS \cite{kazakh_tts}, Yodas \cite{yodas}. These datasets provide diverse recording conditions and speaking styles, contributing to domain robustness.

\subsubsection{Crowdsourcing and Quality Control}

We collected Russian, Kyrgyz, Kazakh, and Uzbek speech using an in-house crowdsourcing platform. Each utterance was transcribed by five independent annotators. Quality was monitored with interleaved golden-set items and automatic pruning of ambiguous gold prompts based on inter-annotator character error rate (CER). Final transcriptions were produced using a reliability-weighted ROVER algorithm \cite{rover} with text normalization. We removed annotators from the pool if their performance on golden-set items was inconsistent with their agreement with the ROVER aggregate on non-gold tasks.

\subsubsection{Synthetic Data}

To increase coverage for Kazakh and Kyrgyz, we augment fine-tuning with synthetic speech. We sample text prompts from the Kazakh/Kyrgyz portion of mC4 \cite{mc4}, synthesize audio using an in-house multi-speaker TTS system (over 100 voices), and apply ASR-based verification. Specifically, we decode each synthetic utterance with a preliminary ASR checkpoint and retain it only when the CER between the prompt and the ASR hypothesis is below a threshold, filtering out mismatches and low-quality generations.

\subsubsection{Weakly Supervised Data}
We obtain additional supervised-style training data from long-form recordings paired with transcripts via a forced-alignment pipeline. We use MMS \cite{mms} to obtain time-aligned transcripts and derive short training segments. We then apply ASR-based filtering: we decode each segment with a preliminary ASR checkpoint and keep it only if CER between the segment transcript and the ASR hypothesis is below a threshold, discarding likely misaligned or noisy segments.

\section{Experiments}

\subsection{Experimental Setup}

We pre-train the model for 300k steps with a learning rate of $2 \times 10^{-4}$ and a total audio batch size of $2048 \times 32$\,s per update (approximately 18.2 hours of audio). We then fine-tune for 200k steps using AdamW, with a peak learning rate of $6\times10^{-5}$, a 5k-step warmup, and cosine decay to $1\times10^{-7}$. Gradient accumulation is used to achieve a virtual batch size of 3200.

In both stages, data sampling weights are applied at the epoch level by reconstructing the training pool before each epoch according to the target sampling distribution (cluster-level in pre-training; language/domain-level in fine-tuning). Unless stated otherwise, pre-training uses the empirical cluster distribution, whereas fine-tuning uses uniform sampling across languages with domain-stratified sampling within each language.

\subsection{Evaluation\label{evaluation}}

We evaluate on Common Voice \cite{common_voice}, FLEURS \cite{fleurs}, and internal in-the-wild datasets collected via crowdsourcing (6--20\,h of speech-dense audio per language). We exclude utterances longer than 30 seconds and samples whose references contain digits. References are normalized before WER computation (lowercasing, punctuation removal); model outputs are normalized where required to ensure consistency. For English, we additionally canonicalize British/American spelling variants (e.g., ``centre'' $\to$ ``center'') and expand common contractions (e.g., ``I'm'' $\to$ ``I am'') in both references and hypotheses. Whisper large-v3 \cite{whisper} does not support Kyrgyz and is run without a language token. Seamless-M4T v2 large \cite{seamless} is used with explicit language forcing, as it otherwise tends to translate into a more frequent language. Omnilingual ASR \cite{omnilingual} is the 1B variant with an LLM decoder, with language specified per utterance. All models are decoded using greedy decoding.

\begin{table}[t]
\caption{Pre-training data mixture and cluster sampling weights; values are averaged WER (\%) across available splits per language. E0 uses the natural cluster distribution.}
\label{tab:pretrain_mixture}
\centering
\setlength{\tabcolsep}{4pt}
\resizebox{\linewidth}{!}{%
\begin{tabular}{l c ccccc}
\toprule
\textbf{Exp} & \large $[p_{\text{C}1},p_{\text{C}2},p_{\text{C}3},p_{\text{C}4},p_{\text{C}5}]$ &
\textbf{Rus} & \textbf{Eng} & \textbf{Kir} & \textbf{Kaz} & \textbf{Uzb} \\
\midrule
E0  & [0.60, 0.27, 0.08, 0.03, 0.02] & \textbf{4.6} & \textbf{14.4} & 9.4 & 12.3 & 10.5 \\
E1  & [0.60, 0.20, 0.10, 0.05, 0.05] & \textbf{4.6} & 14.6 & 9.1 & 11.9 & 10.2 \\
E2  & [0.50, 0.15, 0.25, 0.05, 0.05] & 4.7 & 15.4 & \textbf{8.5} & \textbf{11.4} & \textbf{9.7} \\
E3  & [0.40, 0.25, 0.25, 0.05, 0.05] & 4.9  & 14.6 & 8.7 & 11.6 & 9.8   \\
\bottomrule
\end{tabular}}
\end{table}

\subsection{Pretrain Data Mixture}

We ablate cluster-level sampling weights during pre-training to improve WER on Central Asian languages (Table~\ref{tab:pretrain_mixture}). E0 uses the natural (unweighted) distribution; E1 and E2 progressively shift weight toward C3 (Central Asian cluster). From E0 to E2, WER on Kyrgyz improves from 9.4 to 8.5 and on Uzbek from 10.5 to 9.7, while Russian remains almost unchanged (4.6$\to$4.7) and English WER increases only slightly (14.4$\to$15.4). Experiment E3 attempts to balance performance across all languages, recovering English WER to 14.6\%, but compromises the gains on target languages. Since our primary objective is to improve low-resource performance, we adopt E2 for the final model and public release.

\subsection{Fine-tuning Data Ablation \label{data_ablation}}

In this section, we investigate the impact of fine-tuning data sampling strategies on ASR performance. Table~\ref{tab:finetune_weighting} contrasts three approaches: (i) natural distribution sampling (\textit{unbalanced}), (ii) \textit{language-balanced} sampling with uniform language probabilities, and (iii) \textit{domain-aware} sampling, which applies explicit weighting to sub-domains within each language.

While unbalanced sampling minimizes WER for English (the predominant language), it significantly degrades performance on low-resource languages (Kyrgyz, Kazakh, and Uzbek) due to insufficient representation during optimization. Adopting a language-balanced strategy yields substantial gains for tail languages with only a marginal regression in English performance.

Domain-aware sampling delivers the most significant improvements for Kazakh and Kyrgyz, particularly on the Internal split (spontaneous speech). We attribute this to better regulation of domain composition: unlike simple language balancing, domain-aware sampling prevents large synthetic subsets from dominating the training curriculum, thereby improving generalization to spontaneous conditions. In contrast, Uzbek exhibits a distinct pattern where domain-aware sampling offers no clear benefit over language balancing. This is consistent with its data profile: the Uzbek subset is relatively small and lacks the synthetic augmentation that requires domain-weight regulation.

Based on these findings, our final released model combines the E2 pre-trained encoder with domain-aware sampling for fine-tuning. Table~\ref{tab:model_performance} compares our final ASR system against public multilingual ASR.

\begin{table}[t]
\caption{WER (\%) by fine-tuning sampling strategy.}
\label{tab:finetune_weighting}
\centering
\setlength{\tabcolsep}{4pt}
\resizebox{\linewidth}{!}{%
\begin{tabular}{llccc}
\toprule
Lang & Split & \shortstack{unbalanced\\ sampling} & \shortstack{language-balanced\\ sampling} & \shortstack{domain-aware\\ sampling} \\
\midrule
\multirow{2}{*}{English} & CV     & \textbf{19.6} & 21.1 & 21.5 \\
                        & FLEURS & \textbf{8.7}  &  10.8 &  9.4 \\
\midrule
\multirow{3}{*}{Russian} & CV      & 6.0 & 6.2 & \textbf{5.1} \\
                         & FLEURS  & 3.2        & 3.1 & \textbf{3.0} \\
                         & Internal& \textbf{6.0}  & 6.2 & \textbf{6.0} \\
\midrule
\multirow{3}{*}{Kazakh} & CV      & 15.3 & \textbf{13.6} & 13.8 \\
                        & FLEURS  &  4.7 & \textbf{4.4}  & \textbf{4.4} \\
                        & Internal& 18.0 & 17.4          & \textbf{15.8} \\
\midrule
\multirow{3}{*}{Kyrgyz} & CV      & 11.2 & 10.5 & \textbf{10.2} \\
                        & FLEURS  &  5.9 & 5.6  &  \textbf{5.5} \\
                        & Internal& 11.4 & 11.2          & \textbf{9.8} \\
\midrule
\multirow{3}{*}{Uzbek} & CV      & 13.0 & \textbf{8.9}  &  9.2 \\
                       & FLEURS  &  10.4 & \textbf{7.0}  &  7.3 \\
                       & Internal& 15.5 & 12.8 & \textbf{12.7} \\
\bottomrule
\end{tabular}}
\end{table}

\subsection{Encoder Ablation\label{encoder_ablation}}

To isolate the contributions of the pretrained encoder and the fine-tuning data recipe, we fine-tune Whisper Large and Omnilingual-1B encoders with a CTC objective on our dataset, using the same setup as for GigaAM. We also run the same pipeline with our smaller 240M model, in addition to the full 600M model. Table~\ref{tab:encoder_ablation} summarizes the results: both GigaAM variants outperform Whisper and Omnilingual under this controlled setup. Notably, the 240M encoder outperforms both larger baselines across all languages except English, achieving an average WER of 12.2\% versus 14.1\% for Whisper and 16.6\% for Omnilingual, despite being substantially smaller and faster.

\begin{table}[t]
\caption{Encoder ablation under matched CTC fine-tuning: WER (\%) averaged over splits per language. Best in \textbf{bold}, second-best \underline{underlined}}
\label{tab:encoder_ablation}
\centering
\resizebox{\linewidth}{!}{%
\begin{tabular}{lcccc}
\toprule
Lang & \textbf{\shortstack{GigaAM\\240M}} & \textbf{\shortstack{GigaAM\\600M}} & \textbf{\shortstack{Omni SSL 1B\\CTC}} & \textbf{\shortstack{Whisper Large v3\\ CTC}} \\
\midrule
English & 19.1 & \textbf{14.4} & 24.5 & \underline{16.7} \\
Russian           & \underline{6.4} & \textbf{4.6}  & 11.8 &  9.4 \\
Kazakh            & \underline{13.7} & \textbf{12.3} & 19.0 & 17.1 \\
Kyrgyz            & \underline{10.2} & \textbf{9.4}  & 13.6 & 13.4 \\
Uzbek             & \underline{11.7} & \textbf{10.5} & 14.2 & 13.8 \\
\midrule
Avg               & \underline{12.2} & \textbf{10.2} & 16.6 & 14.1 \\
\bottomrule
\end{tabular}}
\end{table}

\subsection{Tail languages fine-tuning}

To further assess long-tail transfer, we evaluate adaptation to two languages with minimal pre-training coverage (less than 500 hours): Bashkir (bak) and Georgian (kat), selected from different clusters. We fine-tune GigaAM, Whisper, and Omnilingual encoders under the same CTC setup and train a separate model per language using only its Common Voice training data (93h for Georgian, 143h for Bashkir). GigaAM achieves the lowest WER on both languages (Table~\ref{tab:unseen_lang}).

\begin{table}[ht]
\caption{Adaptation to minimal-coverage tail languages: WER (\%) on Common Voice (separate per-language CTC model)}
\label{tab:unseen_lang}
\centering
\begin{tabular}{l c c}
\toprule
\textbf{Model} & \textbf{Georgian} & \textbf{Bashkir} \\
\midrule
Whisper Large v3 CTC &13.4 & 11.1 \\ 
Omnilingual SSL 1B CTC &7.8  &8.2 \\
GigaAM & \textbf{3.8} &\textbf{3.6} \\
\bottomrule
\end{tabular}
\end{table}

\section{Conclusion}
This study addresses the challenge of developing high-quality ASR for underrepresented languages under severe data imbalance. We introduce GigaAM Multilingual, a Conformer encoder pre-trained on 2M hours of audio using a cluster-aware sampling strategy. Our experiments demonstrate that explicit data balancing during pre-training and domain-aware sampling during fine-tuning are critical to prevent head-language dominance. As a result, our model significantly outperforms strong open baselines on Kazakh, Kyrgyz, and Uzbek, particularly in spontaneous speech. Notably, under matched CTC fine-tuning, even our compact 240M model surpasses the significantly larger Whisper Large v3 and Omnilingual 1B encoders, highlighting the efficiency of our approach. Beyond ASR performance, the resulting encoder enables rapid adaptation with limited supervision and proves effective for data curation. By releasing our model weights, we provide a robust foundation for future research and production systems in low-resource settings.

\clearpage

\section{Use of Generative AI Disclosure}
Generative AI tools were used for limited language editing and phrasing support during manuscript preparation (e.g., improving clarity, grammar, and wording of selected paragraphs). The authors reviewed and revised all AI-assisted edits and take full responsibility for the content, claims, experimental results, and conclusions of this paper. Generative AI tools were not used to generate experimental results, analyze data, or make scientific decisions.

\bibliographystyle{IEEEtran}
\bibliography{mybib}

\end{document}